\documentclass[conference]{IEEEtran}
\IEEEoverridecommandlockouts
\usepackage{cite}

\usepackage{amsmath,amssymb,amsfonts}
\usepackage{array}
\usepackage{graphicx}
\usepackage{textcomp}
\usepackage{xcolor}
\def\BibTeX{{\rm B\kern-.05em{\sc i\kern-.025em b}\kern-.08em
    T\kern-.1667em\lower.7ex\hbox{E}\kern-.125emX}}
\usepackage{algorithm} 
\usepackage{algpseudocode} 
\usepackage{amsmath}
\usepackage{float} 
\usepackage{multirow}
\usepackage{subfigure}

\usepackage{makecell}
\usepackage[export]{adjustbox} 

\begin{document}

\title{Optimising Intrusion Detection Systems in Cloud-Edge Continuum with Knowledge Distillation for Privacy-Preserving and Efficient Communication}
\author{\IEEEauthorblockN{1\textsuperscript{st} Soad Almabdy}
\IEEEauthorblockA{\textit{School of Computing, Engineering \& the Built Environment} \\
\textit{Edinburgh Napier University}\\
Edinburgh, UK}
\IEEEauthorblockA{\textit{Faculty of Computing and Information Technology} \\
\textit{King Abdulaziz University}\\
Jeddah, Saudi Arabia\\
soad.almabdy@napier.ac.uk}
\and
\IEEEauthorblockN{2\textsuperscript{nd} Amjad Ullah}
\IEEEauthorblockA{\textit{School of Computing, Engineering \& the Built Environment} \\
\textit{Edinburgh Napier University}\\
Edinburgh, UK \\
a.ullah@napier.ac.uk}
}
\maketitle
\begin{abstract}
The growth of the Internet of Things has amplified the need for secure data interactions in cloud-edge ecosystems, where sensitive information is constantly processed across various system layers. Intrusion detection systems are commonly used to protect such environments from malicious attacks. Recently, Federated Learning has emerged as an effective solution for implementing intrusion detection systems, owing to its decentralised architecture that avoids sharing raw data with a central server, thereby enhancing data privacy. Despite its benefits, Federated Learning faces criticism for high communication overhead from frequent model updates, especially in large-scale Cloud-Edge infrastructures. This paper explores Knowledge Distillation to reduce communication overhead in Cloud-Edge intrusion detection while preserving accuracy and data privacy. Experiments show significant improvements over state-of-the-art methods.   
\end{abstract}


\section{Introduction}\label{sec:introduction}
The Cloud-Edge continuum enables data processing across cloud and edge layers, spanning from centralised servers to edge devices~\cite{yu2022high}. It facilitates seamless data flow, ensuring efficient handling of large-scale datasets to meet the growing demand for faster interactions. As networks grow in complexity, traditional security measures fall short. To address these, Intrusion Detection Systems (IDS) are commonly employed within distributed environments to detect and mitigate potential attacks~\cite{roy2022lightweight}. Recently, machine learning (ML) and deep learning (DL) have been widely used in IDS development due to their high effectiveness~\cite{yu2022high}. However, they face challenges such as data privacy, communication overhead, and high computational demands~\cite{roy2022lightweight}. Moreover, their reliance on centralised data collection raises privacy concerns and limits data transfer. 

Federated Learning (FL) enables decentralised model training by keeping data on distributed devices, eliminating the need for centralisation. Local models are trained on devices and aggregated into a global model, making FL a promising approach for IDS in Cloud-Edge ecosystems. However, FL suffers from high communication overhead, especially with frequent model updates in large networks. Factors like server traffic, packet loss, bandwidth constraints, and device heterogeneity further degrade performance. This paper, alongside recent research \cite{liu2020deep,qin2020line,zhao2022semisupervised,simra2024enhancing}, aims to address these challenges.

This paper adopts a two-fold approach to tackle communication overhead and computational constraints in FL-based IDS. First, we introduce edge servers as intermediaries to improve efficiency in the client-server model. Second, we leverage Knowledge Distillation (KD) for model compression, preserving performance while reducing communication costs on resource-constrained edge devices. We propose a hierarchical FL-based intrusion detection model integrating KD, optimised for Cloud-Edge ecosystems.
\section{Related Works}\label{sec:relatedwork}
Several studies have proposed various types of distributed models for implementing IDS within the Cloud-Edge continuum. For example, Mert et al.~\cite{nakip2023decentralized}  leverages cooperative learning among IDSs deployed in cyber-systems to ensure data confidentiality while allowing systems to share collective experiences and local data. Zhuotao and Chunhua ~\cite{lian2022decentralized} approach utilises neural networks to enhance accuracy and safeguard the integrity of local data by leveraging FL. Han et al. \cite{wang2021non} proposed P2PK-SMOTE for training supervised ML to improve performance with non-IID data. Othmane et al. \cite{friha20232df} introduced the IDS framework to protect industrial smart IoT devices from cyber-attacks. All of these solutions are distributed and focus mainly on resolving a single point of failure. In contrast, this paper focuses mainly on reducing the communication overhead caused by the FL process. 

Several studies have focused specifically on reducing communication overhead. For example, Yi et al.~\cite{liu2020deep} integrated FL with gradient compression for anomaly detection in time-series data, reducing communication costs by 50\%.  However, gradient compression reduces communication load by lowering precision but can slow model convergence~\cite{shah2021model}, potentially requiring more communication rounds to achieve the desired accuracy. Qiaofeng et al.~\cite{qin2020line} uses BNNs in programmable data plane switches for FL-based packet classification, minimising communication overhead even in multi-edge network scenarios. However, binarisation may be not suitable for complex architectures demanding high accuracy~\cite{qin2020binary}.

Several studies particularly explored the potential of KD. For example, Ruijie et al.~\cite{zhao2022semisupervised} integrated KD with semi-supervised FL to improve edge device classification using unlabelled data. Tabassum et al.~\cite{simra2024enhancing} explored IoT security, evaluating FL with KD and DL for IDS enhancement. Their study found that this approach improves threat detection, accelerates learning, and strengthens privacy in IoT systems. Rabaie et al.~\cite{benameur2024novel} also integrated KD with FL to improve computational efficiency. These solutions are closely related to ours, however, integrate KD with traditional FL-based client-server (centralised) architectures. While KD reduces communication overhead while preserving accuracy, frequent global model updates remain costly, particularly in large, dynamic environments requiring real-time learning~\cite{lin2020ensemble}. In contrast, this paper addresses communication overhead by integrating KD with a hierarchical FL approach, optimised for the layered Cloud-Edge architecture.
\section{Proposed Framework: Synergising KD with FL}\label{sec:framework}
\begin{figure}[t]
\centering
\fontsize{8pt}{10pt}
\includegraphics[width=0.4\textwidth]{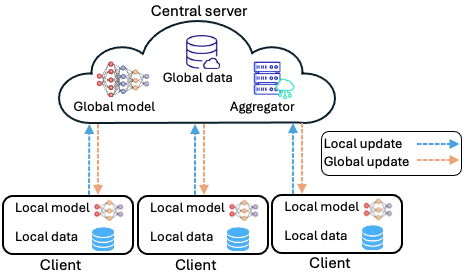}
\caption{Traditional (Centralised) Federated Learning}\label{fig:cfl} 
\end{figure}

The traditional FL architecture (depicted in Fig.~\ref{fig:cfl}) consists of a two-layer client-server model, where a single central server aggregates model updates from several clients.  In a Cloud-Edge ecosystem, such a centralised approach requires high-bandwidth transmission, causing significant communication overhead. To address this issue, several model compression techniques, for example, BNNs, KD, and gradient compression are employed in existing research works~\cite{liu2020deep,qin2020line, benameur2024novel} to reduce the size of the data sent. However, frequent updates from clients can still add up the communication overhead in large-scale Cloud-Edge networks.

This paper explores a hierarchical FL-based architecture with intermediaries that aggregate client updates and manage communication between clients and the central server, reducing data transfer and conserving bandwidth. This structure is well-suited for the Cloud-Edge ecosystem, where computation is distributed across cloud and edge layers. Leveraging this hierarchy, we integrate KD, enabling knowledge transfer from larger models to smaller, more efficient ones. This allows resource-constrained edge devices to benefit from complex models while significantly reducing communication overhead, enhancing the overall efficiency of FL-based IDS.  

\begin{figure}[t]
\centering
\fontsize{8pt}{10pt}
\includegraphics[width=0.48\textwidth]{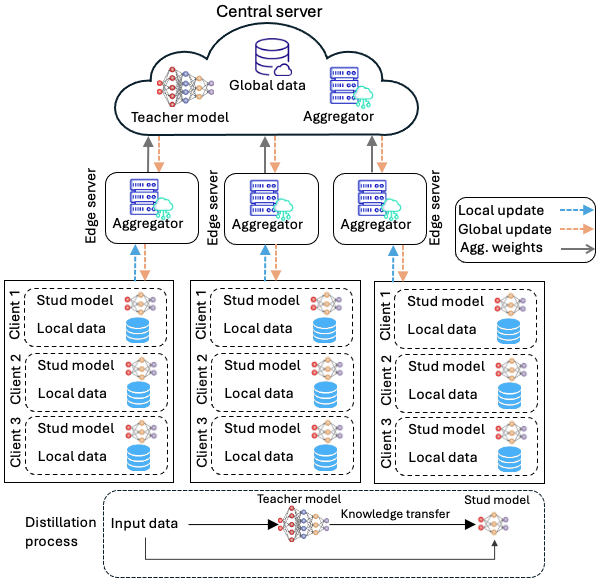}
\caption{Proposed Framework}\label{fig:proposed_model} 
\end{figure}

Fig.~\ref{fig:proposed_model} presents the architecture of the proposed approach. It comprises three layers---cloud, edge, and device---each hosting key components: the \textit{Central server}, \textit{Edge servers}, and \textit{Clients}, respectively. The \textit{Central server}---runs the teacher model, trained on global data---aggregates client weights updates the global model, and distributes weights to edge servers. The \textit{Edge servers} mediate between \textit{Clients} and the \textit{Central server}, aggregating client weights and forwarding them upstream. They also distribute global model updates to \textit{Clients}. Lastly, \textit{Clients} train a lightweight student model using local data, perform KD, and communicate with edge servers for model updates. 

The algorithm~\ref{algFL_IDS} further illustrates the overall working mechanism of the proposed approach. The process begins with initialisation, where the central server sets up the teacher model (\(TM_{\text{global}}\)) using a public dataset. Clients, grouped into \(k\) clusters, each run a student model (\(M_{\text{student}}\)) under an edge server. During each communication round \(r = 1, 2, \ldots, R\), the teacher model distills knowledge to student models. Clients train their models using local data (\(D_{\text{private}}\)) and send updated weights (\(W_{\text{local}}\)) to their edge server. In local aggregation, edge servers combine client weights using Federated Averaging (Eq~\ref{eq:local_aggregation}) and forward them to the central server.

For global aggregation, the central server updates the teacher model using Eq~\ref{eq:global_aggregation} and distributes the updated model to edge servers, who further send it to clients. This process repeats across rounds, with distillation loss (\(Loss_{\text{Distill}}\)) guiding student models to align with the teacher model (Eq~\ref{eq:distillation_loss}), where \(\alpha\) balances direct learning from labels and knowledge transfer. After \(R\) rounds, the model's performance is evaluated.

\begin{align}
\begin{split}\label{eq:local_aggregation}
    W_{\text{local}} = \frac{\sum_{i=1}^{n} W_{client}^{\text{i}}}{n}
\end{split}\\ 
\begin{split}\label{eq:global_aggregation}
   W_{\text{Global}} = \frac{\sum_{i=1}^{k} W_{local}^{\text{i}}}{k}
\end{split}\\ 
\begin{split}\label{eq:distillation_loss}
  Loss_{\text{Distill}} = \alpha \cdot Loss_{\text{Hard}} + (1 - \alpha) \cdot Loss_{\text{Soft}}
\end{split}    
\end{align}

\begin{algorithm}[t]
\footnotesize 
\caption{FL-based IDS with Knowledge Distillation}
\label{algFL_IDS}
\textbf{Inputs:}
\begin{itemize}
    \item $TM_{global}$: Global Teacher Model at the central server 
    \item $M_{student}$: Student models at clients 
    \item $D_{private }^i$: Private dataset of client $i$ 
    \item $R$: Number of global communication rounds  
    \item $N$: Number of clients 
    \item $K$: Number of clusters
\end{itemize}

\begin{algorithmic}[1]
\Procedure{\textnormal{FederatedKD}}{$TM_\text{global}, M_\text{student}$}
    \State \textbf{Initialise} ${TM}_{global}$ using public dataset
    \For{each communication round $r = 1$ to $R$}
        \For{each cluster $k = 1$ to $K$}
        \State Distill knowledge from $TM_{global}$ to $M_{student}$
            \For{each client $i$ in cluster $k$}
                \State Minimise combined loss:
                \State  $\mathrm{Loss} = \alpha \cdot \text{Loss}_{Soft} + (1 - \alpha) \cdot \text{Loss}_{Hard}$
                \State Train $M_{student}$ using private dataset $D_{private}^i$
                \State Send local weights $W_{client}$ to edge server
            \EndFor
            \State Perform  aggregation at edge server using FedAvg:
            \State \hspace{1em} $W_{local} = \frac{1}{|N|} \sum_{i=1}^N W_{client}^{i}$
        \EndFor
        \State Send aggregated updates to central server
        \State Perform  aggregation at central server using FedAvg:
        \State \hspace{1em} $W_{global} = \frac{1}{K} \sum_{i=1}^K W_{local}^i$
        \State Redistribute global update to edge servers
        \State Edge server redistributes global update to each clients
    \EndFor
    \State \textbf{Evaluate} the model after $R$ communication rounds
\EndProcedure
\end{algorithmic}
\end{algorithm}

For implementation, both teacher and student models use CNNs. The teacher model, a larger CNN trained on a public dataset, enables advanced feature extraction. The student model, a smaller CNN with fewer layers and parameters, runs on resource-constrained edge devices. Using KD, it inherits the teacher’s performance while maintaining efficiency and reducing computational cost. Table~\ref{table:teacher_student_model} lists the specifics details of the models.
\begin{table}[t]
\centering
\small 
\caption{Comparison of Teacher and Student Model Structures}
\label{table:teacher_student_model}
\setlength{\tabcolsep}{0.4em} 
\renewcommand{\arraystretch}{1.2} 

\begin{tabular}{|p{2.2cm}<{\raggedright}|p{1.3cm}<{\centering}|p{1.3cm}<{\centering}|p{1.0cm}<{\centering}|p{1.0cm}<{\centering}|}  
\hline
\multirow{2}{*}{\textbf{Layer Type}} & \multicolumn{2}{c|}{\textbf{Model Output Shape}} & \multicolumn{2}{c|}{\textbf{Parameters}} \\ \cline{2-5}
                                     & \textbf{Teacher} & \textbf{Student} & \textbf{Teacher} & \textbf{Student} \\ \hline
Conv1D             & (28, 32)   & (28, 16)   & 128       & 64       \\ \hline
MaxPooling1D       & (14, 32)   & (14, 16)   & 0         & 0        \\ \hline
Conv1D             & (12, 64)   & (12, 32)   & 6,208     & 1,600    \\ \hline
MaxPooling1D       & (6, 64)    & (6, 32)    & 0         & 0        \\ \hline
Flatten            & 384        & 192        & 0         & 0        \\ \hline
Dense (ReLU)       & 128        & 64         & 49,408    & 12,352   \\ \hline
Dense (Softmax)    & 6          & 6          & 774       & 390      \\ \hline
\multicolumn{3}{|l|}{\textbf{Total Parameters}} & \textbf{56,518} & \textbf{14,406} \\ \hline
\multicolumn{3}{|l|}{\textbf{Trainable Parameters}} & \textbf{56,518} & \textbf{14,406} \\ \hline
\end{tabular}

\end{table}

\section{Evaluation} \label{sec:evaluation}
Our evaluation focuses on three key objectives. First, we assess the overall performance of the proposed approach using accuracy, precision, recall, and F1 score. Second, we analyse its communication efficiency and its impact on performance. Finally, we compare its effectiveness with two closely related state-of-the-art methods.
\subsection{Dataset and experimental setup}\label{sec:datasetAndExperimental}
We evaluate our approach using the Edge-IIoTset dataset~\cite{ferrag2022edge}, which includes six classes: one normal and five attack types—(D)DoS, Information Gathering, Man-in-the-Middle (MITM), Injection, and Malware. The dataset consists of 63 features and 2,219,201 records, divided into 12 shards: shards 1–9 for clients and 10–12 are used as public datasets.

Additionally, the dataset was preprocessed to enhance model training and testing. The steps included: (1) removing records with missing or inaccurate values, (2) discarding irrelevant flow features such as ports, IP addresses, payload data, and timestamps, (3) standardising data by subtracting the mean and scaling to unit variance, (4) encoding categorical features into numerical arrays, and (5) normalising features to a 0–1 range to improve model convergence.

Lastly, for the experiments, we utilised nine clients, grouped into three clusters, each managed by an edge server, with a single global central server. The knowledge distillation settings included a temperature of 5, a distillation loss weight (\(\alpha\)) of 0.5, a learning rate of 0.001, a batch size of 32, and two local training epochs per round, optimised using the Adam optimiser.

\subsection{Results and Discussion} \label{sec:results}
\subsubsection{Performance evaluation} \label{sec:resPerformance}
\begin{figure}[b]
\centering
\includegraphics[width=0.9\linewidth]{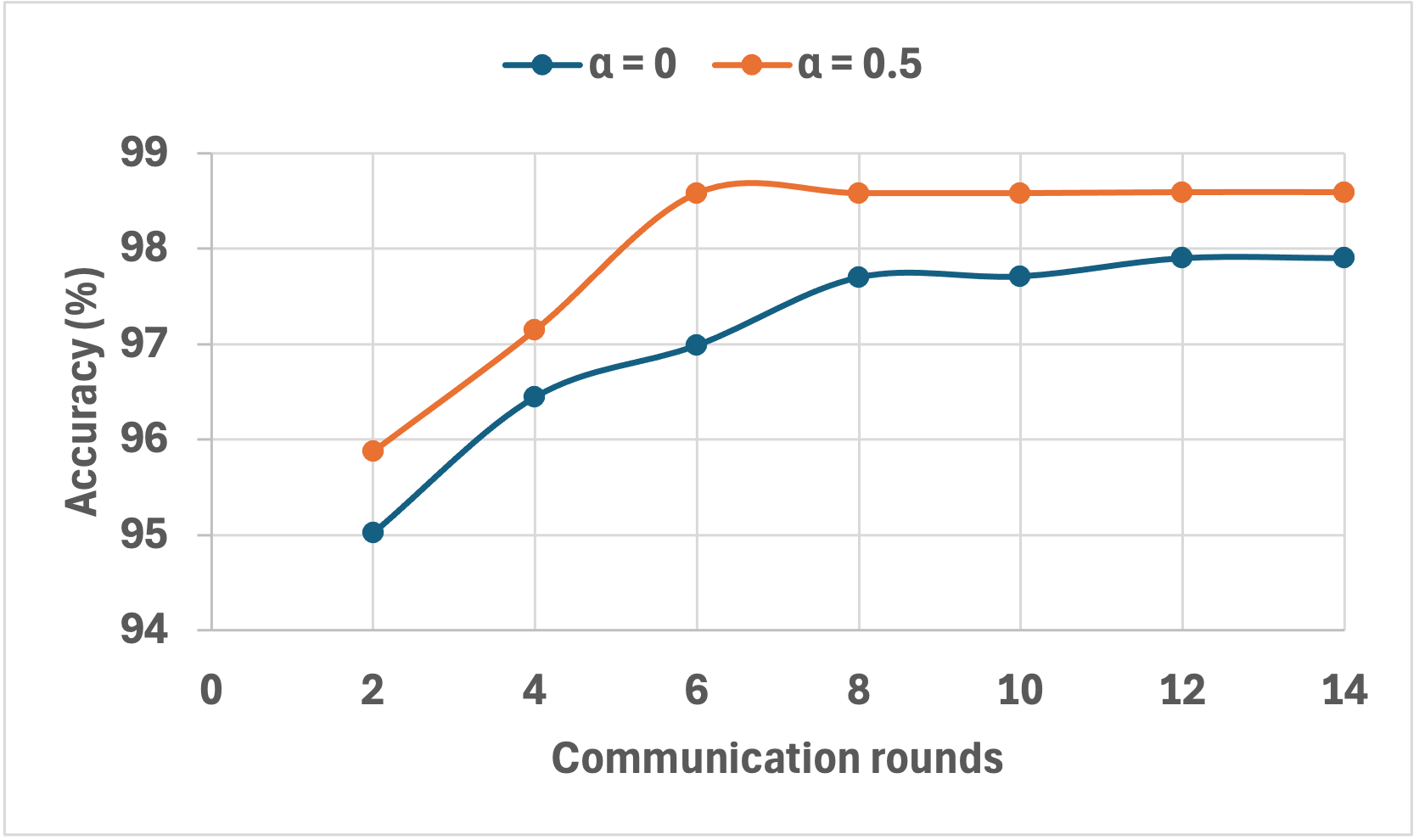}
\caption{Impact of KD on overall performance across communication rounds}\label{fig:Round} 
\end{figure}

Table~\ref{table:Attack} provides the overall performance, as well as a detailed breakdown for each attack type, while Table~\ref{tab:fl_ids_performance} shows the performance in communication rounds (2, 4, 6, 8, 10). As expected, the model initially showed lower overall performance as the model relied on local data and server updates. Gradually, as the number of communication rounds increased, the accuracy improved, reaching 98.58\% at round 6, indicating convergence. Additionally, we also evaluated the impact of the KD process on the overall performance of the proposed approach. More specifically, we considered the following two scenarios:
\begin{enumerate}
    \item The proposed model runs without KD mechanics. In this case, the distillation weighting factor \(\alpha\) is zero.
    \item The proposed model operates using KD mechanics, with the \(\alpha\) = 0.5. 
\end{enumerate}
Table~\ref{table:Per_Eval} summarises the results, showing that with KD (i.e. when \(\alpha\) = 0.5), the model outperforms in all aspects. Notably, KD not only improves overall performance but also accelerates model training. As shown in Figure ~\ref{fig:Round}, without KD (\(\alpha\) = 0.0), the highest accuracy of 97.90\% is reached after 12 rounds, while with KD (\(\alpha\) = 0.5), 98.58\% accuracy is achieved in just six rounds. The findings show that KD with hierarchical FL-based IDS accelerates local model training on the client side.
\begin{table}[t]
\centering
\fontsize{8pt}{10pt}
\caption{Overall performance of the proposed model}
\label{table:Attack}
\setlength{\tabcolsep}{0.6em}
\renewcommand{\arraystretch}{1.1}%
\begin{tabular}{|l|l|c|c|c|c|c|}
\hline
 \multicolumn{2}{|l|}{}&\textbf{Accuracy} & \textbf{Precision} & \textbf{Recall} & \textbf{F1-score} \\
\hline
\multicolumn{2}{|l|}{\textbf{Overall}}& 98.58 & 0.9864 & 0.9858 & 0.9854 \\
\hline

\multirow{6}{*}{\rotatebox[origin=c]{90}{\textbf{Attack types}}}&Normal & 100 & 1.00 & 1.00 & 1.00 \\  \cline{2-6}
&MINM & 100 & 1.00 & 1.00 & 1.00 \\ \cline{2-6}
&Malware & 94 & 1.00 & 0.94 & 0.97 \\ \cline{2-6}
&Injection & 86 & 0.99 & 0.86 & 0.92 \\ \cline{2-6}
&DoS/DDoS & 100 & 0.93 & 1.00 & 0.96 \\ \cline{2-6}
&Inf. gathering & 83 & 0.94 & 0.88 & 0.94 \\ 
\hline
\end{tabular}
\end{table}
\begin{table}[t]
\centering
\fontsize{8pt}{10pt}
\caption{Overall performance of the proposed model across communication rounds}
\label{tab:fl_ids_performance}
\setlength{\tabcolsep}{1em}
\renewcommand{\arraystretch}{1.2}%
\begin{tabular}{|c |c |c |c |c|}
\hline
\textbf{Round} & \textbf{Accuracy} & \textbf{Precision} & \textbf{Recall} & \textbf{F1-score} \\
\hline
2  & 95.88 & 0.9500 & 0.9500 & 0.9301 \\ \hline
4  & 97.15 & 0.9721 & 0.9715 & 0.9710 \\ \hline
6  & 98.58 & 0.9864 & 0.9858 & 0.9854 \\ \hline
8  & 98.58 & 0.9864 & 0.9858 & 0.9854 \\ \hline
10 & 98.58 & 0.9864 & 0.9858 & 0.9854 \\ \hline

\end{tabular}
\end{table}
\begin{table}[b]
\centering
\fontsize{8pt}{10pt}
\caption{Impact of KD on overall performance}
\label{table:Per_Eval}
\setlength{\tabcolsep}{0.6em}
\renewcommand{\arraystretch}{1.2}%
\begin{tabular}{|c|c|c|c|c|}
\hline
\textbf{Distillation loss weight} & \textbf{Accuracy} & \textbf{Precision} & \textbf{Recall} & \textbf{F1-score} \\ \hline
\textbf{\(\alpha\) = 0.0} & 97.90 & 0.9800 & 0.9790 & 0.9783 \\ \hline
\textbf{\(\alpha\) = 0.5} & 98.58 & 0.9864 & 0.9858 & 0.9854 \\ \hline
\end{tabular}
\end{table}
\subsubsection{Communication Efficiency}
The communication efficiency of the proposed approach was assessed using metrics, including transmission time, aggregation time, data size, memory usage, and inference time. First, we compare the proposed hierarchical approach with the centralised approach. Second, we assess the impact of the KD with weighting factors (\(\alpha\)=0) and (\(\alpha\)=0.5). This resulted in four scenarios, i.e. with and without KD for both architectures. 
\paragraph{Transmission time}
Measuring transmission time per round helps estimate communication overhead and model update delays. Table~\ref{table:tranmission_time} presents the results obtained in summarised form for each of the four scenarios. In the table, the 'C to S' represents the average transmission time from Clients (C) to the Central server (S), and 'S to C' the reverse. In the case of the proposed approach, the 'C to S' includes transmission from \textit{Clients} to the \textit{Edge Servers} and then to the \textit{Central Server} and 'S to C' as the reverse. This table shows that the proposed KD approach achieves the lowest average transmission time, highlighting significant efficiency improvements in the communication process.

\paragraph{Aggregation time} Table ~\ref{table:tranmission_time} also presents the aggregation time for all four scenarios. It is evident from this table that the proposed approach with KD achieved a significantly lower aggregation time of 0.06 milliseconds, resulting in a faster and more efficient overall process.

\begin{table}[t]
\centering
\caption{Transmission time. Clients (C), Central Server (S)}
\label{table:tranmission_time}
\setlength{\tabcolsep}{0.6em}
\renewcommand{\arraystretch}{1.2}%
\begin{tabular}{|l|l|c|c|c|}
\hline
\multicolumn{2}{|l|}{\textbf{Scenarios}} & \textbf{C to S} & \textbf{S to C} & \textbf{Agg time}\\ \hline

\multirow{2}{*}{\textbf{Centralised FL}}&\textbf{\(\alpha\) = 0.0} & 3.91 & 3.76 & 0.18  \\ \cline{2-5}
&\textbf{\(\alpha\) = 0.5} & 3.59 & 3.47 & 0.11 \\ \hline

\multirow{2}{*}{\textbf{Proposed}}&\textbf{\(\alpha\) = 0.0} & 3.54 & 3.5131 & 0.14 \\ \cline{2-5}
&\textbf{\(\alpha\) = 0.5} & \textbf{3.31} & \textbf{3.30} & \textbf{0.06} \\ \hline
\end{tabular}
\end{table}

\begin{figure*}[t]
    \centering
    \subfigure[Inference time]{\includegraphics[width=0.28\textwidth]{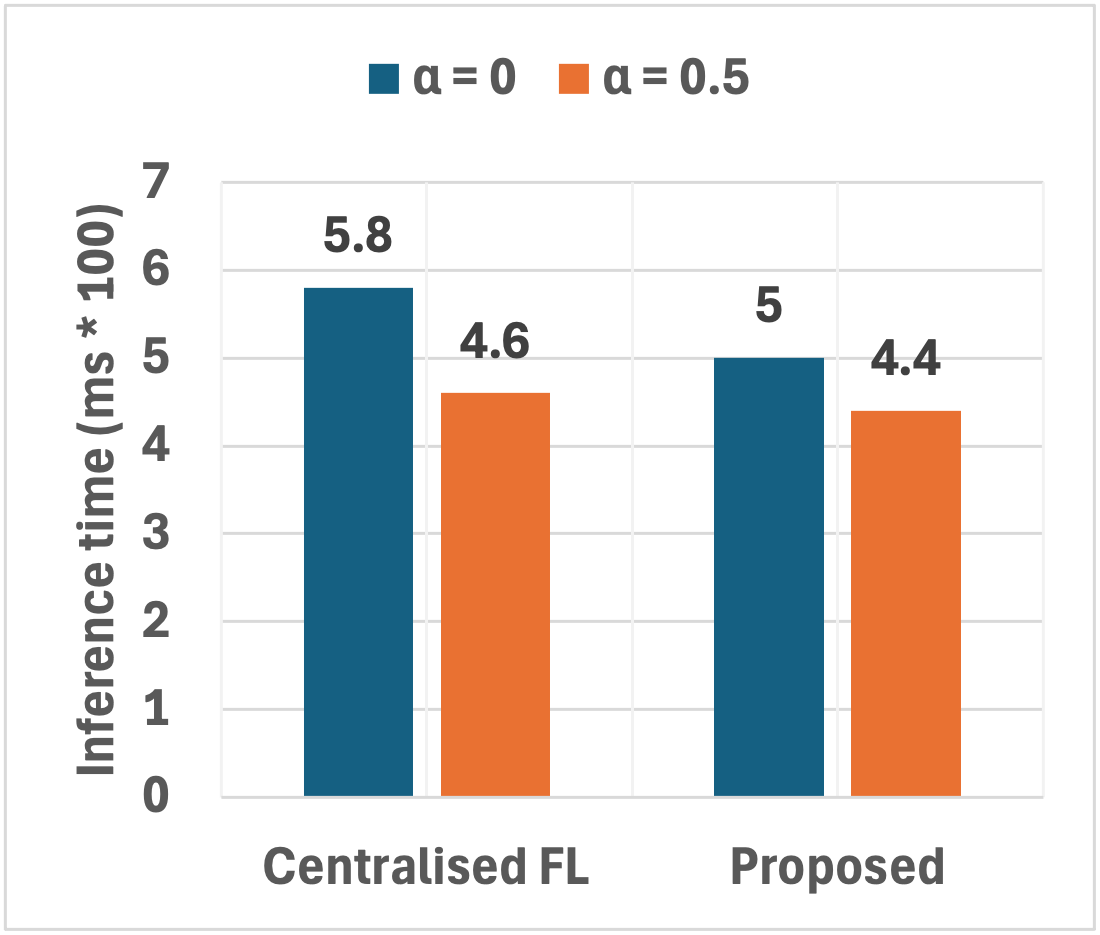}
    \label{fig:inference}} 
    \hspace{0.0cm}
    \subfigure[Memory]{\includegraphics[width=0.28\textwidth]{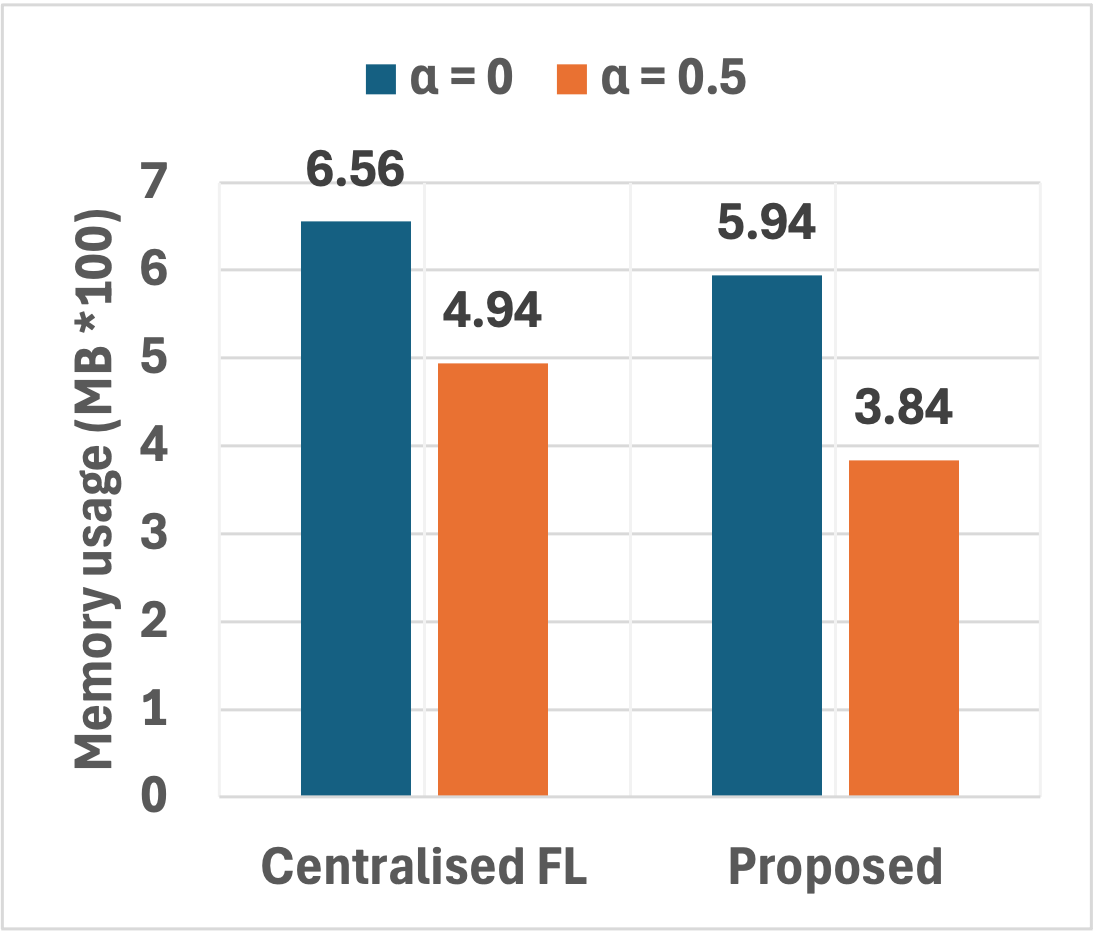}
    \label{fig:memory}} 
    \hspace{0.0cm}
    \subfigure[Data size]{\includegraphics[width=0.28\textwidth]{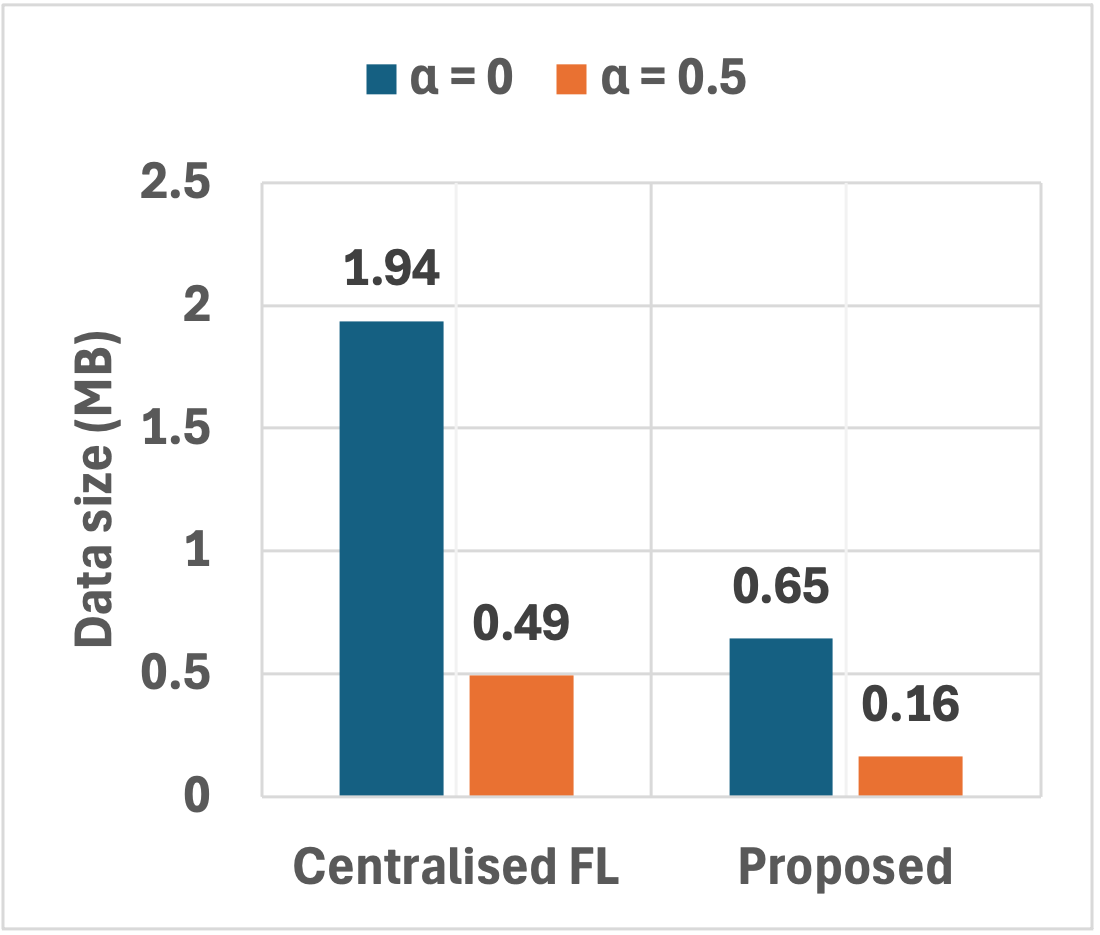}
    \label{fig:datasize}}
    \caption{Other results}
    \label{fig:results}
\end{figure*}

\subsubsection{Inference time }
The inference time measures the model's speed in making predictions on unseen data. Figure~\ref{fig:inference} demonstrates that our approach with KD achieves the lowest inference time amongst all scenarios. Using a compact KD model reduces computational burden, enabling faster inference. Even slight improvements are vital for real-time intrusion detection, enhancing reliability in dynamic environments.
\subsubsection{Memory usage }
Figure~\ref{fig:memory} presents memory usage across four scenarios, highlighting the higher consumption in centralised FL due to managing individual client updates. In contrast, the hierarchical approach reduces memory usage by aggregating client updates at edge servers before forwarding compacted updates to the \textit{Central Server}. By transmitting a smaller data volume, as depicted in (Figure~\ref{fig:datasize}), the proposed method reduces memory demands at each server level, improving efficiency.
\subsubsection{Data size}
Figure~\ref{fig:datasize} illustrates the impact of data size for each scenario, highlighting that the proposed hierarchical approach significantly reduces data size in comparison to the centralised FL model from 1.94 MB to 0.65 MB in the case without KD, achieving a reduction of 74.51\% and from 0.49 MB to 0.16 MB in the case of KD, achieving a reduction of 66.67\%. This highlights the effectiveness of the hierarchical structure, especially with KD, in minimizing data size, reducing network and central server load, enhancing scalability, and supporting resource-constrained environments.
\subsection{Proposed model vs SOTA}
Lastly, we compared the performance of our proposed model against two of the most recent SOTA models, including Tabassum et al.  \cite{simra2024enhancing} and Rabaie et al. \cite{benameur2024novel}. Both these research works are closely related, i.e. both are CNN-based and proposed the integration of KD with the FL approach as ours, however, using the traditional client-server architecture. Table~\ref{table:SOTA} reports the summarised results. It is clear from this table that the overall performance of the proposed approach surpasses that of the other two models. Moreover, the inference time for the proposed model is significantly lower, at just 0.44 seconds, compared to 163 seconds reported by Rabaie et al.~\cite{benameur2024novel}. Lastly, it is important to note that, in comparison with Rabaie et al.~\cite{benameur2024novel}, who achieved an accuracy of 84.55\% at round 30, our model attained 98.58\% accuracy in just 6 rounds.

\begin{table}[t]
\centering
\caption{Proposed model vs SOTA}
\label{table:SOTA}
\setlength{\tabcolsep}{0.6em}
\renewcommand{\arraystretch}{1.1}%
\begin{tabular}{|c|c|c|c|c|}
\hline
\textbf{Model} & \textbf{Accuracy} & \textbf{F1-score} & \textbf{Inference Time} & \textbf{Rounds} \\ \hline
\textbf{\cite{simra2024enhancing}} & 94.90\% & 94.80 & - & - \\ \hline
\textbf{\cite{benameur2024novel}} & 84.55\% & - & 163.00 s & 30 \\ \hline
\textbf{Proposed} & 98.58\% & 98.54 & 0.44 s & 6 \\ \hline
\end{tabular}

\end{table}

\section{Conclusion}~\label{sec:conclusion}
This paper proposes integrating knowledge distillation with federated learning in a hierarchical architecture for intrusion detection in Cloud-Edge ecosystems. The approach reduces communication overhead while maintaining privacy and high detection accuracy. Our evaluation shows that knowledge distillation significantly cuts communication overhead without sacrificing performance. By training smaller student models, clients transfer fewer parameter updates, reducing communication needs. This also speeds up local model training, enabling faster convergence with fewer communication rounds. Future work will involve developing a realistic test-bed to validate the approach and explore challenges in real-world deployments.

\bibliographystyle{IEEEtran}

\end{document}